\documentclass[sigconf]{acmart}

\usepackage{booktabs} 
\usepackage[utf8x]{inputenc}
\usepackage{amsmath}
\usepackage[linesnumbered,ruled]{algorithm2e}
\usepackage{latexsym}
\setcopyright{rightsretained}

\acmDOI{10.475/123_4}

\acmISBN{123-4567-24-567/08/06}

\copyrightyear{2017}
\acmYear{2017}
\setcopyright{acmlicensed}
\acmConference{WI '17}{August 23-26, 2017}{Leipzig, Germany}\acmPrice{15.00}\acmDOI{10.1145/3106426.3106498}
\acmISBN{978-1-4503-4951-2/17/08}

\begin{document}
\title{Concept-aware Geographic Information Retrieval}

\author{Noemi Mauro}
\affiliation{%
  \institution{University of Turin}
  \streetaddress{Computer Science Dept.}
  \city{Torino} 
  \state{Italy} 
  \postcode{10149}
}
\email{noemi.mauro@unito.it}

\author{Liliana Ardissono}
\affiliation{%
  \institution{University of Turin}
  \streetaddress{Computer Science Dept.}
  \city{Torino} 
  \state{Italy} 
  \postcode{10149}
}
\email{liliana.ardissono@unito.it}

\author{Adriano Savoca}
\affiliation{%
  \institution{University of Turin}
  \streetaddress{Computer Science Dept.}
  \city{Torino} 
  \state{Italy} 
  \postcode{10149}
}
\email{savoca@di.unito.it}

\renewcommand{\shortauthors}{N. Mauro et al.}

\begin{abstract}
Textual queries are largely employed in information retrieval to let users specify search goals in a natural way. However, differences in user and system terminologies can challenge the identification of the user's information needs, and thus the generation of relevant results. 
We argue that the explicit management of ontological knowledge, and of the meaning of concepts (by integrating linguistic and encyclopaedic knowledge in the system ontology), can improve the analysis of search queries, because it enables a flexible identification of the topics the user is searching for, regardless of the adopted vocabulary. 

This paper proposes an information retrieval support model based on semantic concept identification. Starting from the recognition of the ontology concepts that the search query refers to, this model exploits the qualifiers specified in the query to select information items on the basis of possibly fine-grained features. Moreover, it supports query expansion and reformulation by suggesting the exploration of semantically similar concepts, as well as of concepts related to those referred in the query through thematic relations. A test on a data-set collected using the OnToMap Participatory GIS has shown that this approach provides accurate results.
\end{abstract}




\maketitle

\begin{figure*}
\includegraphics[width=2.1\columnwidth]{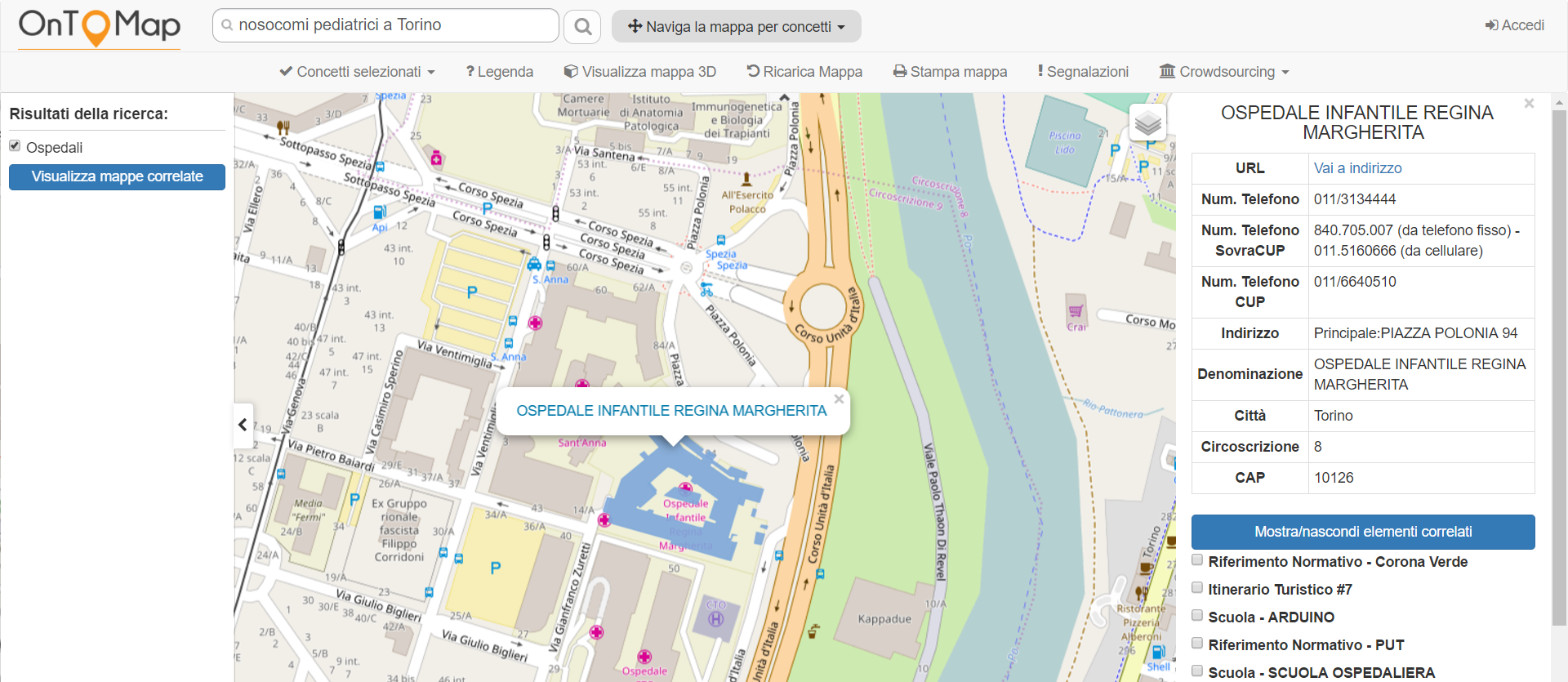}
\caption{Search results for ``nosocomi pediatrici (children clinics) a Torino" and visualization of the data concerning a specific hospital.}
\label{f:pediatrici}
\end{figure*}

\section{Introduction}
Finding information in large datasets can be challenging, without a support that helps understand what can be looked for. With respect to pure category-based search, textual queries are a fairly natural interaction mean. However, differences between the user's and system's domain conceptualizations can compromise the identification of the user's information needs, and thus the provision of appropriate results.

We argue that, in the interpretation of textual queries, the integration of semantic and linguistic knowledge can improve the system's capability to provide relevant results because:
\begin{itemize}
\item  
It makes it possible to deal with queries expressed in different terminologies (e.g., by taking synonyms and word similarity into account), abstracting from the domain conceptualization adopted by the system, that the user is probably unaware of.
\item  
It supports an explicit identification of the concepts on which the user focuses, preventing misunderstandings.
\item
It enables the expansion of queries with thematically related concepts, thus broadening the scope of the search results, depending on the user's interests.
\end{itemize}
Both aspects contribute to overcoming the limitations of pure keyword-based search, which can fail to retrieve the desired data due to word mismatch, or that can return irrelevant results because it lacks word disambiguation.

Focusing on Web-GIS, which are the topic of this work, we developed an interactive query interpretation model that jointly uses linguistic, encyclopaedic, and an ontological representation of domain knowledge to answer geographical queries.
Our approach follows the associative information retrieval model \cite{Giuliano-Jones:62} but is based on the execution of two query interpretation phases: 
\begin{enumerate}
\item  
Semantic concept identification, by matching a semantically expanded query to the domain ontology in order to identify the referenced concepts. This enables the retrieval of a set of information items belonging to the general topics of the search query; e.g., hospitals. 
\item  
Facet-based filtering of results to take the qualifiers specified in the query into account; e.g., {\em pediatric} hospitals. Also in this case, the semantics of qualifiers is taken into account to abstract from the terminology used by the user. 
\end{enumerate}
This two-steps approach supports the generation of relevant results because information is filtered on a semantic basis. Assuming a correct identification of the concepts referenced in the query, results cannot include items belonging to concepts different from those directly or indirectly expressed by the user. Moreover, this approach supports query reformulation and expansion, e.g., by relaxing the qualifiers, or by exploiting the semantic relations defined in the ontology in order to select more general, or thematically related, concepts than the one specified in the original queries.

This paper presents our model and describes how it is applied to support information search in the OnToMap Participatory GIS \cite{Voghera-etal:16,OnToMap}, which supports information sharing and participatory decision-making.
A test on a dataset collected within the OnToMap project revealed that this approach provides accurate results.

This work builds on the preliminary work presented in \cite{Ardissono-etal:16}, which sketched the query interpretation model described here, and extends it with the interpretation of textual queries including qualifiers, and with the presentation of preliminary test results.

The remainder of this paper is organized as follows: Section \ref{related} positions our work in the related one. Section \ref{ontomap} provides an overview of the OnToMap application. Section \ref{model} describes our query interpretation model. Section \ref{experiments} describes the results of a preliminary evaluation of our approach and Section \ref{conclusions} concludes the paper and outlines our future work.

\section{Background and Related Work}
\label{related}

A flexible interpretation of textual queries presupposes that the system is able to map them to its own domain conceptualization. This mapping is particularly difficult because, as discussed in \cite{Belkin:80}, information retrieval occurs in an anomalous state of knowledge: basically, in a search task the user is asked to specify something that (s)he does not know. Indeed, it is very likely that her/his terminology differs from the one of the system and the two have to be reconciled to identify the user’s information needs. 

Query expansion techniques have been long explored to enhance information retrieval. For instance, \cite{Qiu-etal:93} proposed a statistical approach to the selection of  terms for query expansion, based on the analysis of the whole query (instead of single words) and on development of a custom thesaurus inferred from the source pool of documents.  Moreover, \cite{Mandala-etal:99} showed that the integration of different types of thesauri (linguistic, domain specific, etc.) improves the performance of query expansion techniques with respect to the adoption of individual ones. \cite{Grootjem-vanDerWeide:06} suggests to create local thesauri, tailored to the query and to the collection being searched, and proposes a conceptual query expansion based on the combination of terms that are meaningful for the collection and form a ``formal concept". Finally, \cite{Berg-Schuemie:99} proposes to exploit Self-Organizing Maps to automatically generate associative conceptual spaces based on word co-occurrence in document spaces, saving the effort to build ad-hoc thesauri.
With respect to these works, we do not attempt to define new algorithms for word sense disambiguation, but a new way to combine external services for query interpretation. Our model exploits the linguistic functions offered by sophisticated external word disambiguation services for query expansion. However, taking into account the difficulties in expanding short queries, it enhances the flexibility of concept recognition by enriching the domain ontology with linguistic and encyclopaedic knowledge that makes it possible to associate further synonyms and keywords to concepts. Thus, the expanded queries can be matched to a larger, but controlled, set of terms, relevant to the application domain. Moreover, if the system identifies multiple concepts, it proposes them to the user and asks her/him to select the interesting ones for continuing the information search task. As the identified concepts are semantically related to the query, this disambiguation phase is an opportunity to discover related concepts, and other portions of the information space to be explored.

Several GIS use ontologies for conceptualizing the domain \cite{Fonseca-etal:00} and helping users in information retrieval. For instance, SIAPAD \cite{Molina-Bayarri:11} combines semantic knowledge representation with task-based information to map the keywords occurring in search queries to the ontology concepts related to the corresponding activities. With respect to that work, we adopt a general approach, based on linguistic and encyclopaedic knowledge, in order to make the system independent of the execution of particular tasks, which would require the representation of task-specific knowledge. Moreover, the multi-faceted conceptual domain representation used by OnToMap makes it possible to search for information under different points of view.

Some systems support multi-faceted information browsing, but this is not related to textual query interpretation. For instance, \cite{Stadler-etal:16} presents a graphical user interface for faceted exploration of geographical Linked Data, but the navigation of the information space is done by browsing a set of hierarchical menus, with the possibility of specifying search keywords. In comparison, OnToMap supports both graph-based exploration, based on the visualization of views on the domain ontology, and a textual one, which directly maps natural language queries to ontology concepts.

Other GIS, such as TripAdvisor \cite{TripAdvisor}, ask for a separate specification of geographical entities and information to be found. They use the keywords included in the query to match geo-data names, item reviews, etc., providing mixed results that include heterogeneous items (e.g., items tagged by the keyword, or having it in their own names, addresses, etc.). Similarly, OpenStreetMap \cite{OpenStreetMap} applies keyword-based search offered by Nominatim and returns all the items located in the bounding box that include the specified tags and keywords. 
MapQuest \cite{Mapquest} supports looking for three types of information: place, address and categories. The category-based search is similar to the one offered by TripAdvisor. MapQuest offers an extended set of categories corresponding to information layers, that can be added or removed from the map.
Different from all these systems, OnToMap identifies the concepts referenced in the query to retrieve coherent results, e.g., all the sport facilities located in the selected geographical area. Moreover, it supports Linked Data exploration based on the semantic relations among ontology concepts.

Wikimapia \cite{Wikimapia} supports category-based search by presenting a list of categories that users can browse, with an auto-completion search bar. Categories reflect the tags that users insert when they add new crowdsourced items to the map, and tags can be organized in an hierarchical structure. In comparison, OnToMap offers a textual interaction mode, and an ontology-based {\em navigation by concepts}, for semantically browsing both subclass and thematic relations between concepts.

Some recent work on information filtering attempts to acquire relations among information types from the observation of users' behaviour, and is complementary to our work. For instance, Google search engine manages the Knowledge Graph \cite{GooglekNowledgeGraph} to relate facts, concepts and entities depending on their co-occurrence in queries. On a related perspective, CoSeNa \cite{Candan-etal:09} employs keyword co-occurrence in the corpus of documents to be retrieved, and ontological knowledge about the domain concepts, to support the exploration of text collections using a keywords-by-concepts graph. The graph “supports navigations using domain-specific concepts as well as keywords that are characterizing the text corpus”.

Finally, recent search auto-completion models, such as COMMA \cite{Porrini-etal:14}, support the search of items in catalogs by indexing information items and by applying string-matching algorithms for item selection. Our work differs in two main aspects: firstly, we rely on item classification in ontology concepts to reduce the amount of pre-processing work to be done by the system. Secondly, we exploit domain-dependent and linguistic knowledge about ontology concepts, as well as word sense disambiguation, to support query interpretation by abstracting from the terminology used by the user.

\section{Overview of OnToMap}
\label{ontomap}

OnToMap supports the management of interactive community maps for information sharing and participatory decision-making \cite{Voghera-etal:16}. 
It enables both the consultation of spatial data and the creation of public and private geographical maps, which reflect individual information needs and can be enriched with crowdsourced content to help project design and group collaboration.

\subsection{Information Search Support}
\label{search}

OnToMap offers two information search modes, both based on a semantic representation of domain knowledge, which is formalized as an ontology specifying the main concepts and relations that characterize the information space: 
\begin{itemize}
\item  
In the {\em navigation by concept} mode, out of the scope of this paper, the user browses a graph depicting the ontology concepts, and (s)he can select the relevant ones to visualize the corresponding items in the maps. 
\item  
In the {\em textual mode}, the user can submit textual queries formulated in her/his own vocabulary. The system attempts to match the words occurring in the queries to the ontology concepts, possibly suggesting query expansions to help the user find the needed information, or visualize other related results (you might be also interested in ...).
\end{itemize}

While interacting with OnToMap, the user can specify textual queries that include a geographical reference, or (s)he can combine queries with the selection of an area in the map. 
Regardless of the interaction mode, OnToMap displays the results on a map focused on the geographical area delimited by the identified bounding box. However, the user can dynamically change the bounds (via zoom and drag actions) to view results belonging to different areas. 

The background layer of the map is based on OpenStreetMap \cite{OpenStreetMap} to present a rich picture of the selected geographical area. On top of this, the information items resulting from the search query are highlighted; they are displayed using vivid colours (or pointers). 
For instance, Figure \ref{f:pediatrici} displays the geometry of hospital ``Ospedale Infantile Regina Margherita" in blue, and is zoomed on the main hospital area of Torino. The item is the singleton result of a query searching for the pediatric hospitals in the town. Around it, there are other hospitals (see the cross icons), which are not highlighted because they are not for children. 

The semantic knowledge representation underlying data retrieval and visualization helps the exploration of the information space in several ways. For instance, the table in the right portion of Figure \ref{f:pediatrici} shows the details (properties) of the ``Regina Margherita" hospital, which the user has visualized by clicking on the item in the map. Moreover, by clicking on button ``Mostra/Nascondi elementi correlati" (show/hide related items), the user can visualize other information, related to the item in focus via semantic and geographic relations. For example, the right portion of Figure \ref{f:pediatrici} provides links to a school (``Arduino") and to some official documents on land usage concerning the area of the geographical item (``Riferimento normativo" - normative reference).

\subsection{Knowledge Representation}

The domain conceptualization underlying OnToMap is based on an OWL ontology \cite{W3C-OWL} supporting:
\begin{itemize}
\item
A multi-faceted specification of the concepts and relations characterizing the information space, which is structured on the basis of different high-level perspectives (natural, artificial and landscape plan), specialized into more detailed concepts; see \cite{Ardissono-etal:16}.
\item
The integration of heterogeneous data \cite{Fonseca:02} and their management as Linked Data \cite{Janowicz-etal:12}. We used the ontology to integrate Open Data from the Municipality of Torino, Metropolitan Torino City, and Piedmont Region.
\item 
Graph-based information exploration; see Section \ref{search}.
\end{itemize}
OnToMap stores geographical information in a triple store that maintains data in RDF \cite{W3C-RDF} format. 
The triple store is queried via GeoSPARQL \cite{GeoSPARQL} queries, generated starting from the ontology concepts selected by analyzing the search query. The result of a query is a GeoJSON FeatureList \cite{GeoJSON}, i.e., a list of GeoJSON objects, each one representing a different information item, whose attributes are mapped to the properties defined in the ontology concepts.

\section{Semantic Interpretation of Textual Queries}
\label{model}

For the definition of our textual query interpretation model, we analysed a dataset of geographical queries extracted from AOL query log (about 15000 queries).
This helped us recognize a number of typical patterns of the queries, and in particular of their geographical references; see \cite{Ardissono-etal:16}.
The pattern we extrapolated can be represented as follows:
$$\{\{address\}, \{C_1, ..., C_m,\}, \{Q_1, ..., Q_n\}\}$$ where 
\begin{itemize} 
\item 
$\{address\}$ is the geographical reference, defined by bounding box or by geographical entity specification;
\item 
$\{C_1, ..., C_m,\}$ are the (lemmas of) concepts and of synonyms of concepts referenced by the query;
\item 
$\{Q_1, ..., Q_n\}$ are the qualifiers, which characterize the items that the user is looking for within the set of instances of the recognized concepts. Qualifiers derive from the attributes or propositional phrases of the search query.
\end{itemize}
For instance, sentence ``Public schools and transportation in Torino" can be represented as \{\{Torino\}, \{school, transportation\}, \{public\}\}.\footnote{We omit the synonyms of the terms of the query for brevity.} 

The following subsections describes how, starting from a textual query, the above representation is generated. Notice that we adopt a lightweight approach, based on the incremental recognition of query components, possibly relying on external services specialized in the recognition of different components, and their usage to progressively filter the set of data to be returned as a result.


\subsection{Linguistic and Encyclopaedic Knowledge about Concepts}
\label{representation}

As described in \cite{Ardissono-etal:16}, to support a flexible matching between the terminology used in the search queries and the concepts defined in OnToMap, the ontology concepts are enriched with linguistic knowledge that makes their meaning explicit.  
We consider multiple ways to refer to the same concept, through synonyms, as well as linguistic definitions, including largely used descriptions - especially when the concepts are technical. These are a source of relevant keywords to refer to the same concept in natural language expressions. For instance, a possible definition of ``Ospedale” (hospital) is 
``building devoted to healing and assisting ill and injured people". This definition makes the concept relevant to queries referring not only to hospitals, but also to buildings, curing, assisting, ill and injured people.

Each concept {\em C} of the ontology has the following features:
\begin{itemize}
\item  
The lemma of a word {\em w} associated to {\em C} and the lemmas of the synonyms of {\em w}. For instance, the lemma of ``ospedali” is ``ospedale” (same word, but singular). Moreover, the lemmas of its synonyms are “clinica” (clinic), ``nosocomio" (another way to define a hospital, that we translate as clinic), etc..
\item
The lemmas of the keywords belonging to the definition(s) of {\em C}. In the previous example, they include, ``cura"(healing), ``ammalato” (ill), ``ferito” (injured), and others.
\end{itemize}
In order to annotate the ontology with linguistic definitions and synonyms, as well as for Word Sense Disambiguation, we used BabelFy multilingual Entity Linking and Word Sense Disambiguation service \cite{babelfy}, in combination with Morph-it! lemmatizer \cite{Zanchetta-Baroni:05}.

\subsection{Phase 1: Query Pre-processing}
\label{preProcessing}
OnToMap pre-processes the input query and generates a {\em normalized query} to be used for concept identification and filtering of information items. The pre-processing task is carried out as follows (see \cite{Ardissono-etal:16} for details):
\begin{enumerate}
\item  
First, the system identifies the geographical specifications included in the query (if any) and submits them to an external geocoder for resolution. Then, it removes them, because they do not need any further processing, and returns a {\em simplified query}, and the identified bounding box to be used for data retrieval. 
\item
Then, the system submits the simplified query to a word sense disambiguation service to retrieve query-dependent synonyms and splits the simplified query into individual words through stop-word removal.
The system returns the {\em normalized query}, which includes the lemmas of each word retained from the original query and of its synonyms. Also in this case, we used Morph-it! to identify the lemmas of words and Babelfy for word sense disambiguation.
\end{enumerate}
For instance, given query ``nosocomi pediatrici a Torino" (pediatric clinics in Torino), the simplified query is ``nosocomi pediatrici" (pediatric clinics) and the normalized one is \{nosocomio, ospedale, pediatrico, infantile\} (\{hospital, clinic, pediatric, paediatric\}).

\begin{figure}
\includegraphics[width=1\columnwidth]{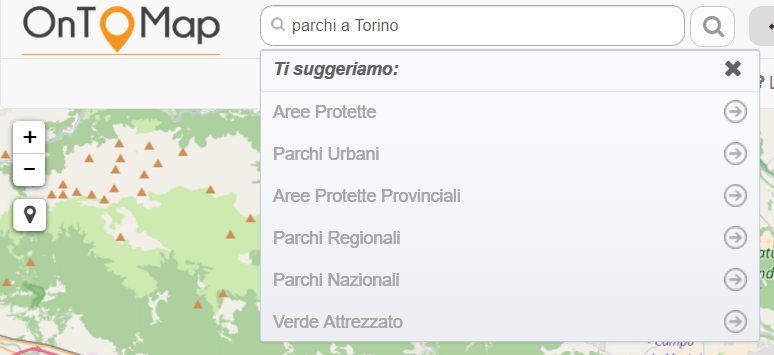}
\caption{Selection of concepts the user is interested in.}
\label{f:disambiguazione}
\end{figure}

\subsection{Phase 2: Concept Identification}
\label{conceptRecognition}

In this phase, the system attempts to match the lemmas of the normalized query to the ontology in order to identify one or more referenced concepts. For any identified concept, it removes the corresponding lemmas (and the lemmas of synonyms) from the normalized query, because they have been resolved. The results of this phase are a set of concepts to be used for data retrieval, and a {\em qualifier set} that only includes the lemmas of the qualifiers (and synonyms), if any.
For instance, given \{nosocomio, ospedale, pediatrico, infantile\}, in this phase the ``ospedale" concept is identified and the set of lemmas associated to the concept are removed from the normalized query. The qualifier set is thus \{pediatrico, infantile\}.

The lemmas of the normalized query can match the ontology concepts in a more or less strict way:
\begin{enumerate}
\item
Direct match between one or more lemmas of the normalized query and those of the ontology concepts, or of their synonyms. In order to find the most specific concepts relevant to the query, concepts are identified by considering single lemmas of the search query as well as adjacent tuples of lemmas. For instance, if the query includes the lemmas of ``public” and ``service”, and the ontology includes both ``services” and a sub-concept ``public services”, the latter concept is identified as a match. 
\item
Match between the lemmas of the normalized query and those of the keywords of the ontology concepts.
\end{enumerate}
If there is a direct match, we assume that the system has successfully interpreted the query and we move to the data retrieval phase (Section \ref{retrieval}). Otherwise, in order to avoid to retrieve irrelevant information, the system attempts to first disambiguate the interpretation by interacting with the user. In this case, it proposes the list concepts it has identified and asks the user to select the interesting ones. Notice that the proposed concepts are related to the words used in the search query through linguistic descriptions and encyclopaedic knowledge. Therefore, they may include concepts more or less loosely related to the user's query, but potentially interesting for expanding the focus of the query and exploring nearby regions of the information space. This phase is thus an opportunity for the user to discover further interesting information.

Figure \ref{f:disambiguazione} shows the disambiguation phase during the interpretation of query ``parchi a Torino" (parks in Torino), that matches the keywords of multiple ontology concepts; e.g., urban, provincial and regional parks, and some types of protected areas. All these concepts are listed in a menu (``Ti suggeriamo: " - we recommend). Given the user's choice, the system moves to the data retrieval phase using the selected concepts for retrieving their instances.

\subsection{Phase 3: Data Retrieval}
\label{retrieval}

Given the set of concepts identified during the previous phase, OnToMap queries the triple store that manages the geographic information to retrieve all their instances located in the specified bounding box.

If the qualifier set is empty (i.e., the normalized query did not contain any qualifiers), this data set represents the result of the search query and the system visualizes it in the map.
Otherwise, a further analysis phase is needed to select the data items whose descriptions are similar to the specified qualifiers; see the next subsection.

\begin{figure}
\includegraphics[width=1\columnwidth]{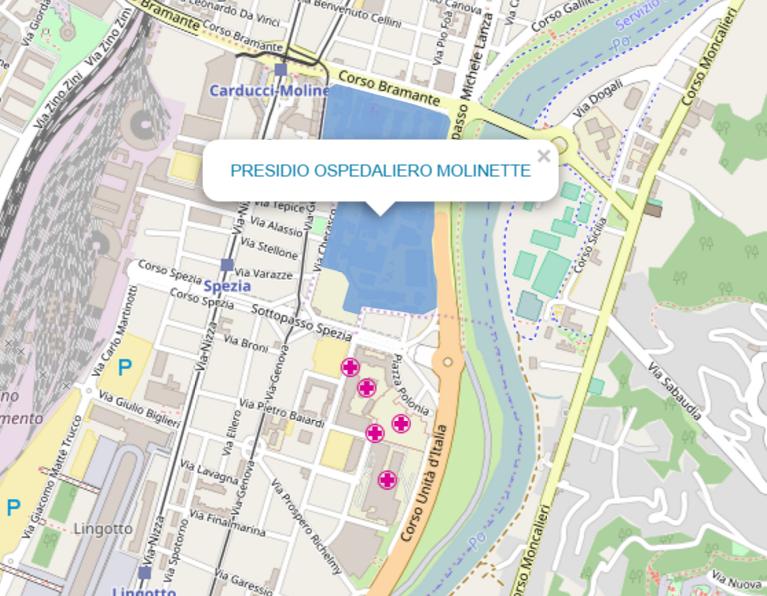}
\caption{Search results for ``Ospedale San Giovanni Battista a Torino".}
\label{f:molinette}
\end{figure}

\begin{algorithm}
 \SetKwInOut{Input}{Input}
 \SetKwInOut{Output}{Output}
 \Input{$p$}
 \Output{true or false}
 $lm = 0$\;
 $similarTerms = 0$\;
\ForEach{($q : Q$)}{
	$lm=\beta * min(|terms(p)|, |terms(q)|)$\;
    \ForEach{($t_p \in terms(p)$)}{
        \ForEach{($t_q \in terms(q)$)}{
        		 $diff=\gamma * max(length(t_p),length(t_q))$\;
          	\If{($LevenshteinDistance(t_p, t_q)\leq diff$)}{
          		$similarTerms++$;
         	 }
        }
        }
        \If{($similarTerms > lm$)}{
			\Return true;
		}
}
	\Return false;
\caption{Checking the Similarity between a Property of an Item and a Qualifier Set}
\label{algoritmo}
\end{algorithm}

\subsection{Filtering Retrieved Data by Qualifiers}
\label{qualifiers}

In this phase, the set of data items selected from the identified concepts is filtered on the basis of the qualifiers occurring in the search query. In fact, the data retrieval phase returns a set of items that could be loosely related to the user's information goals; e.g., all the hospitals in the selected geographical area. However, some of them might not answer the user's requirements. For instance, in our example, starting from the whole list of hospitals in the town, the system should identify those that are pediatric. 

Items are thus analysed to check whether the available information about them is similar to the qualifier set of the query. Also in this case, the user might express her/himself using different words with respect to those occurring in the descriptions of items. However, the inclusion of synonyms during the pre-processing phase guarantees a flexible match between qualifier set and the item information.   
The following points are worth making:
\begin{itemize}
\item
The qualifier set can be matched to item descriptions in a flexible way, without requiring a strict keyword-based correspondence. For instance, in Figure \ref{f:pediatrici}, the system selects ``Ospedale Infantile Regina Margherita" because its name contains attribute ``infantile", that is a synonym of ``pediatrico". 
\item
For the evaluation of the match, all the properties $p$ of the item are considered (e.g., name, typology, etc.), because any of them could bring useful information. For example, in the case of ``Regina Margherita" hospital, the matching information is located in the name. Differently, the name of the hospital visualized in Figure \ref{f:molinette} is ``Presidio Ospedaliero Molinette", but it matches query ``Ospedale Giovanni Battista a Torino" (Giovanni Battista hospital in Torino) because the name of the company managing the hospital, stored in a different property of the item, is ``AOU San Giovanni Battista". 
\end{itemize}
The similarity between an item $i$ and qualifier set $Q$
is evaluated by checking the properties of $i$ against the specifications in $Q$. If at least one property $p$ is similar to at least one specification $q$ of $Q$, $i$ is considered similar to $Q$ and it is retained for visualization in the map. Otherwise, $i$ is filtered out of the result set. 

Algorithm \ref{algoritmo} defines how the similarity between an item property $p$ and a qualifier set $Q$ is evaluated. For each element $q$ of $Q$,\footnote{Both $p$ and $q$ can be composed of more than one term.} 
the algorithm compares each term of $q$ with the terms of $p$ and it counts the number of matching terms ($similarTerms$). If $similarTerms$ is over a threshold ($lm$), $p$ is considered similar to $Q$ and the algorithm returns true. Otherwise, it returns false.

Notice that the retrieved data set can include a large number of items. Therefore, for scalability reasons, the item descriptions cannot be lemmatized. Therefore, the similarity between the terms of $p$ and those of $q$ is evaluated by applying a string metric technique that measures the difference between two words: the Levenshtein Distance \cite{Levenshtein:66}.
$LevenshteinDistance(t_p, t_q)$ is the minimum number of single-character edits (i.e., insertions, deletions or substitutions) required to change $t_p$ into $t_q$. \\For example, 
LevenshteinDistance(ospedale, ospedali) = 1.

In the algorithm, the following notation is used:
\begin{itemize}
\item
$Q$ is the qualifier set of the query.
\item 
$q \in Q$ is a qualifier belonging to $Q$ and can be composed of one or more terms (lemmas); e.g., \{pediatric\}. The terms of $q$ are denoted as $terms(q)$.
\item 
$p$ is a property of $item$ and can be composed of one or more words: $terms(p)$.
\item 
$lm$ is a threshold on the number of similar terms that $p$ and $q$ must include to consider them similar to each other. In order to take into account the fact that qualifiers and properties may include different numbers of terms, $lm$ is computed as a fraction ($\beta$) of the minimum between $|terms(p)|$ and $|terms(q)|$. For our experiments, we set $\beta = 0.5$ to require at least 50\% of similar words between $p$ and $q$, tuned on the length of the set of terms ($p$ or $q$) having minimum cardinality.
\item 
Considering $t_p \in terms(p)$, and $t_q \in terms(q)$, $diff$ is a threshold on the maximum Levenshtein Distance between $t_p$ and $t_q$. As the terms may have different lengths, $diff$ is computed as a fraction $\gamma$ of the maximum length between $t_p$ and $t_q$. For our experiments, we set $\gamma$ to 0.20 to require about 80\% similarity between terms.
\end{itemize}

\section{Preliminary Evaluation of our Query Interpretation Model}
\label{experiments}

\begin{table*}
\centering
\begin{tabular}{|l|c|c|c|c|c|c|}
\hline
\textbf{Type of Queries} & \textbf{\begin{tabular}[c]{@{}c@{}}N°  Queries\end{tabular}} & \textbf{Precision} & \textbf{Recall} & \textbf{F1} & \textbf{\begin{tabular}[c]{@{}c@{}}Std. dev. Precision\end{tabular}} & \textbf{\begin{tabular}[c]{@{}c@{}}Std. dev. Recall\end{tabular}} \\ \hline
Only concepts            & 274                                                            & 1,00               & 1,00            & 1,00        & 0,00                                                           & 0,00                                                        \\ \hline
Concepts + Qualifiers          & 122                                                            & 0,69               & 0,94            & 0,80        & 0,35                                                           & 0,20                                                        \\ \hline
All queries              & 396                                                            & 0,90               & 0,98            & 0,94        & 0,25                                                           & 0,14                                                        \\ \hline
\end{tabular}
\caption{Average accuracy of OnToMap results}
\label{t:average}
\end{table*}

\begin{table*}
\centering
\begin{tabular}{|l|c|c|c|c|}
\hline
\textbf{Concept}      & \textbf{N° of Queries} & \textbf{Precision} & \textbf{Recall} & \textbf{F1} \\ \hline
Hospitals             & 85                     & 0,93               & 0,99            & 0,96        \\ \hline
Schools               & 63                     & 0,78               & 0,93            & 0,85        \\ \hline
Accommodations        & 40                     & 1,00               & 1,00            & 1,00        \\ \hline
Museums               & 29                     & 0,73               & 0,96            & 0,83        \\ \hline
Sport Areas           & 27                     & 0,99               & 1,00            & 1,00        \\ \hline
Places of Worship     & 17                     & 0,57               & 0,88            & 0,69        \\ \hline
Public transportation & 16                     & 1,00               & 1,00            & 1,00        \\ \hline
Public Security       & 14                     & 1,00               & 1,00            & 1,00        \\ \hline
Bus stops             & 12                     & 1,00               & 1,00            & 1,00        \\ \hline
Cinemas               & 9                      & 1,00               & 1,00            & 1,00        \\ \hline
Libraries             & 7                      & 1,00               & 1,00            & 1,00        \\ \hline
Child Care Centres    & 5                      & 0,77               & 1,00            & 0,87        \\ \hline
Street Markets        & 5                      & 0,55               & 1,00            & 0,71        \\ \hline
\end{tabular}
\caption{Concept-based accuracy of OnToMap Results - Queries are grouped by referred concepts}
\label{t:queryByConcept}
\end{table*}

\subsection{Dataset}

We evaluated the accuracy of our model by checking it on a query log that we collected from May 2016 to January 2017, in a number of experiments with users. In these experiments, we asked people to create custom maps for the organization of events in the town, or for participating to a simulated public policy making process. For this purpose, they had to find relevant data using the OnToMap textual information search mode. Users were aware to be logged and gave their consent.

The overall log we collected stores information about different types of activity performed by users while interacting with the system; e.g., search queries, creations of geographical information items, annotations of items, and so forth. In order to respect users' privacy, the system collected anonymous events. 

The original log included 492 queries, but we reduced it to 396 after having removed incomplete or uninterpretable sentences and some other queries that focused on types of information that the system does not handle. For instance, some referred to airports, that are not represented in the OnToMap ontology and for which no information has been imported in the system. 
Of the 396 queries we retained, 122 included a qualifier set while the others only specified the reference geographical area and the main concepts to search for.

Starting from the cleaned log (396 queries), we annotated each query with the ontology concepts it referred to and with the relevant qualifiers to be used in order to retrieve the appropriate data from the OnToMap triple store.

The most frequent concepts searched for in the queries are the following ones: Hospitals (85 queries), Schools (63), Accommodations (40), Museums (29), Sport Areas (27), Places of Worship (17), Public transportation (16), Public Security (14), Bus stops (12). Table \ref{t:queryByConcept} shows the list of concepts that received at least 5 queries.

\subsection{Experiment and Results}

Given the query log, we evaluated the accuracy of OnToMap by comparing the results returned by the system with the items of the dataset that match the annotated queries. The idea is that the annotated queries represented the real information needs expressed by users and we checked them against the system's interpretation by comparing the respective sets of items.

As shown in Table \ref{t:average}, the precision achieved by OnToMap in answering the queries is 0.90, and the recall is 0.98 (see row ``All queries"). 
Moreover, the Standard Deviation of precision and recall on the dataset are, respectively, 0.25 and 0.14, which reveal that the deviation from the means is low. This means that, since the precision and recall are high, the system should have good performance in the interpretation of the search queries. 
However, the results achieved considering the queries that included a qualifier portion (``Concepts + Qualifiers" row of the table) have lower precision, while the recall is satisfactory. 
We hypothesize that the system achieved a lower precision for the following reasons:
\begin{enumerate}
\item  
We adopted a loose interpretation of similarity among items. In fact, an item property is considered similar to the qualifier if the information about it includes at least 50\% terms similar to those of the qualifier. Moreover, an item is similar to another one if it has at least one similar property. We will investigate a stricter definition of item similarity to focus results without downgrading recall.
\item 
The string similarity measure used for the experiment in some cases has low performance. E.g., if a user submits the query "Scuole primarie a Torino" ("Primary school in Torino"), the item set retrieved contains all the primary schools but also some private schools ("Scuola paritaria") because the edit distance between "primaria" and "paritaria" is only 2. We will investigate the performance of other similarity measures, e.g., the Jaro-Winkler distance, to see if they can improve the precision of the algorithm. 
\item   
Similar to what has been done in other applications (e.g., in OpenStreetMap), some words are used as synonyms even though they are only related terms (e.g., exhibition and museum). We must evaluate the benefits of this approach against its problems to decide whether to have stricter keyword sets in the ontology. 
\item  
The lack of an entity recognition function while filtering by qualifiers can cause the selection of false positives. E.g., if a user is searching for "Ospedale San Giovanni Bosco", also the hospital "San Giovanni Battista" is retrieved because it contains the words "San Giovanni". We think that using an entity recognition system could mitigate this problem.
Indeed, stricter interpretations of similarity could be given, asking for a better correspondence between item description and qualifier. However, we don't know the impact of this on recall, given that users might remember only some portions of names of items, or they might input partly wrong information. We will consider this trade-off in our future work by tuning parameters $\beta$ and $\gamma$ in our experiments.
\end{enumerate}

In order to have a better picture of the situation, we analysed queries from a content point of view, looking at the performance of the system when focusing on specific types of information.
Table \ref{t:queryByConcept} reports the accuracy achieved by OnToMap in answering the queries that referred to the most popular concepts that users targeted (the table shows data about the concepts having received at least 5 queries). Looking at the results, it is possible to see that, for some concepts, e.g., places of worship, the system performed rather poorly, with 0.57\% precision and 0.88\% recall, while it did well in many other cases. 
We believe that this variability in accuracy could be related to two causes:
on the one hand, a lack of observations; e.g., concept Places of Worship has been targeted only in 17 queries. On the other, a possible need to refine the domain knowledge of the system by modelling this type of information in a more detailed way; e.g., by representing the different types of place of worship whose information is available in the dataset.

Before concluding this section, we would like to point out that, even though the dataset is small, it is the best we could use to evaluate OnToMap so far. We could not find any public online query logs for the Italian language.
Moreover, having contacted some teams managing search engines in order to ask whether they could provide us with a small set of their own logs, we received negative answers from them. Obviously, we aim at enriching our log with further queries, and use a more representative version of it in future evaluation tasks.

\section{Conclusions and Future Work}
\label{conclusions}

We presented a search query interpretation model supporting semantic, multi-faceted information retrieval. The model is based on an ontological representation of domain knowledge and on its integration with linguistic/encyclopaedic information about the domain concepts in order to enhance query expansion. 


We applied this model to the OnToMap Participatory GIS. In a preliminary evaluation, based on the analysis of a corpus of queries collected by the system in a number of user experiments, our approach has achieved good accuracy results. 

Our future work includes various aspects, among which the validation of our query interpretation model in larger datasets and the analysis of the usefulness of different properties of concepts from the viewpoint of information filtering. At the current stage, the system analyses all the properties of items for answering a search query; however, some of them are less useful than others, and could probably be ignored.  
Our future work also includes the development of personal ontologies, inferred by analysing users' search behaviour \cite{Jiang-Tan:09}
and the acquisition of user models reflecting individual information preferences, given the user's interaction with a map \cite{Wilson-etal:10}. Both aspects are aimed at further improving the system's support to data retrieval, filtering and visualization, in order to reduce the information overload on users.

\section{Acknowledgments}
This work is partially funded by project MIMOSA (MultIModal Ontology-driven query system for the heterogeneous data of a SmArtcity, “Progetto di Ateneo Torino\_call2014\_L2\_157”, 2015-17). We thank Angioletta Voghera, Luigi La Riccia, Maurizio Lucenteforte and Gianluca Torta for their work in the OnToMap project.

\bibliographystyle{ACM-Reference-Format}

\end{document}